# Effects of Spin Fluctuation and Disorder on Topological States of Quasi 2D Ferromagnet $Fe_{1/5}CrTe_2$

M. Lamba, P. Saha[#], K. Yadav, N. Kamboj, and S. Patnaik*
School of Physical Sciences, Jawaharlal Nehru University, New Delhi 110067
(*Corresponding author: spatnaik@jnu.ac.in)

## Abstract

We present a thorough magnetization and magneto-transport study of the diluted Fe-intercalated $CrTe_2$ family member, $Fe_{1/5}CrTe_2$, a van der Waals ferromagnet. $Fe_{1/5}CrTe_2$ shows an elevated Curie transition temperature of 182 K in comparison to the $Fe_{1/3}CrTe_2$ composition, indicating the sensitive role of Fe concentration in modulating magnetic exchange interactions within the $CrTe_2$ framework. The saturated magnetization exhibits a quadratic dependence with temperature, indicating the presence of long-wavelength spin fluctuations. Analysis of the temperature dependent resistivity reveals a dominant $T^{3/2}$ contribution over the typical $T^2$ behaviour, signalling substantial coupling between conduction electrons and localized spins. The magnetoresistance shows a linear and non-saturating negative field dependency throughout a wide temperature range below $T_C$, which is compatible with the increasing suppression of spin-disorder dispersion related to ferromagnetic spin fluctuations. A thorough analysis of the anomalous Hall effect (AHE) shows that extrinsic skew-scattering contribution, which is associated to Fe-related disorder, dominates the anomalous Hall response. The systematic separation of intrinsic and extrinsic components reveals that, over a wide temperature range, the intrinsic anomalous Hall conductivity scales linearly with the saturation magnetization, despite the substantial extrinsic dominant background. The linear behaviour of intrinsic anomalous Hall conductivity with magnetization is in line with a long wavelength spin-fluctuation framework, where thermal spin disorder lowers net magnetization without significantly altering the underlying electronic structure. These findings reveal $Fe_{1/5}CrTe_2$ as a newly investigated van der Waals ferromagnet where spin fluctuations and disorder coexist with a well-defined intrinsic Berry-curvature contribution to the Hall response.

[#]Current Address: National Physical Laboratory, New Delhi 110012

## Introduction

Following the identification of intrinsic long-range ferromagnetism in two-dimensional van der Waals materials [1, 2], layered magnetic systems have drawn significant attention due to their potential in low-power spintronic applications [3, 4] and as platforms for investigating low-dimensional magnetism and spin-dependent transport. Among van der Waals magnetic systems, chromium-based layered chalcogenides are a highly adaptable class that allows for the efficient tuning of magnetic anisotropy, carrier density, and magnetic exchange interactions through intercalation or chemical substitution [1, 2, 4-8]. Cr-based van der Waals systems exhibit a wide range of electronic, chemical structure and magnetic characteristics, which can offer perpendicular magnetic anisotropy [9] and tunability of properties [10] by varying the Cr concentration. The Cr-Te binary system, in particular, produces a rich set of self-intercalated compounds that are succinctly characterized by the generic formula $Cr_{1+\delta}Te_2$, where $\delta$ is the fraction of excess Cr atoms that occupy interlayer sites [11-18]. These compounds adopt roughly comparable crystal structures with alternate fully occupied Cr layers and vacancy layers stacked along the c-axis, depending on the Cr concentration and synthesis conditions [11-21]. Interstitial Cr atoms between $CrTe_2$ layers dramatically alter the electronic structure and magnetic exchange interactions, resulting in a wide range of ground states with characteristics ranging from metallic to semiconducting [18, 22, 23]. The majority of compositions exhibit ferromagnetic ordering, with Curie temperatures ranging from roughly 160 K to 360 K [11-23]. This highlights how strongly magnetism is sensitive to structural distortions and Cr concentration [22, 23].

Although the self-intercalation of Cr in $CrTe_2$ has been studied extensively, controlled intercalation of other transition-metal ions into $CrTe_2$ remains poorly investigated. Specifically, Fe intercalation offers additional flexibility to adjust magnetic anisotropy, exchange interactions, and carrier scattering than is possible with Cr stoichiometry alone [24]. One such example is $Fe_{1/3}CrTe_2$ ($FeCr_3Te_6$), which crystallizes in the P-3m1 structure and can be viewed as an intercalated form of $CrTe_2$ in which Fe occupies van der Waals gap sites [24]. This compound has been found to possess a sizable anomalous Hall response, topological Hall effect, and exhibits ferromagnetic ordering close to 120 K [24]. Notably, the Hall behaviour of this system is largely dominated by the extrinsic skew scattering mechanism, which is partly explained by the occupational disorder of Fe in the interlayer area [24]. These findings demonstrate how disorder and scattering significantly influence the magneto-transport characteristics of Fe-containing $CrTe_2$ systems [24].

In this study, we examine $Fe_{1/5}CrTe_2$, a diluted Fe-containing $CrTe_2$ family member. $Fe_{1/5}CrTe_2$ has a higher Curie temperature ($\approx$ 182 K) than $Fe_{1/3}CrTe_2$, suggesting that Fe concentration is important in adjusting magnetic exchange interactions. There are noticeable signatures in the temperature dependence of resistivity and magnetization that are compatible with magnetization controlled by long-wavelength spin fluctuations. The existence of Fe-related disorder and its effect on scattering processes make $Fe_{1/5}CrTe_2$ an attractive platform to study the competition between extrinsic and intrinsic mechanisms in defining the anomalous Hall response.

Ferromagnetic materials generally showcase three different contributions to their Hall response: the anomalous Hall effect due to broken time-reversal symmetry in the presence of spin-orbit coupling [25, 26], the conventional Hall effect due to the Lorentz force [26], and, in some cases, an additional topological Hall contribution related to noncoplanar spin textures [27]. From a microscopic perspective, the AHE comprises an intrinsic contribution controlled by the Berry curvature of occupied electronic states [28, 29] as well as extrinsic components like skew scattering [30, 31] and side-jump mechanisms [32] resulting from impurity scattering. The intrinsic process is independent of impurity scattering and is based on spin–orbit interaction that is inherent in the crystal band structure [28, 29]. Generally, the intrinsic anomalous Hall conductivity (AHC) is not expected to show a straightforward proportionality to magnetization since it is sensitive to the specifics of the electronic structure and the intensity of the spin–orbit interaction In fact, the intrinsic AHC exhibits nonlinear or even non-monotonic behaviour with regard to magnetization in a number of ferromagnets, which reflects the intricate evolution of Berry curvature close to the Fermi level [34, 35].

However, it's intriguing to note that earlier research on $Mn_5Ge_3$ has demonstrated that the intrinsic AHC can scale linearly with magnetization over a wide temperature range [36]. In contrast to the intrinsic Berry-curvature contribution, which frequently produces nonlinear behaviour [34, 35, 37] linear scaling is naturally expected for extrinsic scattering mechanisms making this result noteworthy [36, 38]. The linear relationship that was seen in $Mn_5Ge_3$ was explained in terms of spin-fluctuation, where thermal spin disorder reduces the macroscopic magnetization without significantly altering the electronic band topology that controls the Berry curvature [36]. Nonetheless, an open question remains if such simplified intrinsic scaling can endure in chemically disordered, intercalated van der Waals magnets, where extrinsic mechanisms may predominate and scattering effects are substantial. Intercalated systems often result in large extrinsic Hall contributions by concurrently changing charge transport channels, introducing disorder, and modifying exchange interactions.

Here, we distinguish between the intrinsic and extrinsic components of the anomalous Hall effect via a systematic magneto-transport study of $Fe_{1/5}CrTe_2$. Our study shows that skew-scattering contributions dominate the total Hall response, which is consistent with considerable scattering linked to disorder connected to the Fe sites. However, throughout a broad temperature range, the intrinsic anomalous Hall conductivity shows a strong linear scaling with the saturation magnetization. This scaling pattern is further corroborated by the temperature dependence of magnetization, which exhibits a linear relationship on $T^2$, a property of long wavelength spin fluctuations [36, 39]. Hence, in $Fe_{1/5}CrTe_2$, the intrinsic Hall response is largely controlled by the magnetic order parameter, despite strong extrinsic scattering, as suggested by the combined observation of $M \propto T^2$ behaviour and $\sigma_{int} \propto M$ scaling. These results identify $Fe_{1/5}CrTe_2$ as a novel intercalated van der Waals ferromagnet where significant disorder coexists with Berry-curvature-driven transport, shedding light on the interaction of intrinsic anomalous Hall physics, spin fluctuations, and scattering mechanisms.

**Experimental Techniques**

The single crystals of $Fe_{1/5}CrTe_2$ are synthesized by the conventional solid-state reaction followed by the chemical vapour transport method (CVT). High purity powders of Fe (99.9%), Cr (99.99%) and Te (99.999%) were taken in a stoichiometric ratio and ground for about 30 minutes. The powder was then pressed into 10 mm dia pellet and vacuum sealed ($10^{-3}$ mbar) in a quartz tube. The sample was made to react in a box furnace at 900 °C for 24 hours. After that, the sample is taken out and re-ground for 30 minutes. For the single crystal growth, the polycrystalline sample is vacuum sealed in quartz tube by adding iodine (2 mg/cm$^3$) as a transport agent. The sealed tube is kept in two zone furnace with high temperature zone at 700 °C and low temperature zone at 650 °C for 10 days. The plate shaped shiny single crystals were obtained at the low temperature zone. The crystal phase is confirmed by analyzing the room temperature X-ray diffraction (XRD) pattern obtained through a table top Rigaku Miniflex 600 instrument. A Bruker single crystal X-ray diffraction was used to record the laue pattern at room temperature of a single crystal. The chemical composition of the synthesized sample was identified by using energy dispersive X-ray (EDX) spectroscopy measurements. The magneto-transport measurements were performed using a Cryogenic physical properties measurements system. The electrical resistivity measurements on bulk single crystal of $Fe_{1/5}CrTe_2$ were done by the standard four probe method. To enhance contact quality and keep the contacts from becoming resistive, 25 μm-diameter copper wires were affixed to the electrical contacts using indium soldering. Temperature dependent magnetization and isothermal magnetization were performed by the vibrating sample magnetometer attachment of the physical properties measurements system (PPMS).

## Results and Discussion

### Structural Characterization

Figure 1 presents Rietveld refining of the polycrystalline precursor powder before chemical vapor transport (CVT), inset (i) of Fig. 1 shows the crystal structure and associated Wyckoff locations of the $Fe_{1/5}CrTe_2$ compound which is generated by using VESTA software. As previous report [24] shows the Wyckoff site 1a fully occupied by Cr atoms, while the 1b site is partially occupied by Fe atoms, forming the transition-metal layer. All observed diffraction peaks are refined satisfactory with a goodness-of-fit ($\chi^2$) value of 3.4 within space group P-3m1 (No. 164) [15, 19-21, 24]. The refined lattice parameters are a = b = 3.856 Å, c = 5.987 Å and angles between the crystallographic axes are α = β = 90°, γ = 120°. The room temperature XRD pattern of a single-crystal flake of the $Fe_{1/5}CrTe_2$ sample is displayed in inset (ii), reveals that all detected peaks align with the (00l) Bragg planes, indicating that the single crystal is orientated with the c-axis perpendicular to the flake surface, confirming with layered crystal symmetry [15, 19, 20, 21, 24]. Inset (iii) shows the diffraction spots obtained through the single crystal diffractometer which confirms the single crystalline nature of the synthesized FCT sample. The lattice parameters obtained are a = b = 3.87 Å, c = 6.01 Å and angles between the crystallographic axes are α = β = 90°, γ = 120°. The obtained results are quite well with those derived from Rietveld refinement of the diffraction pattern. These obtained parameters are consistent with the trigonal crystal structure of the sample and also in agreement with the previous report [24]. The intercalation of Fe atom at interstitial sites results increment in the c-lattice parameter and decrement in the a-lattice parameter. The calculated ratio of c/a of analyzed single crystal diffraction pattern is 1.55 which is comparable with the previous report [24]. Due to similar structure factor of Fe and Cr, it is difficult to quantify the site mixing through XRD analysis. Hence, other sensitive techniques such as neutron diffraction is required to study for determining the atomic site distribution.

Insets (iv) and (v) of Fig. 1 shows the scanning electron microscopy (SEM) image of FCT single crystal flake, which reflects layered morphology of the sample. Elemental composition of the synthesized sample obtained through the EDX pattern is $Fe_{0.6}Cr_3Te_6$ ($Fe_{1/5}CrTe_2$). This shows a deficiency of the Fe atoms in the grown sample as compared to the previous report [24]. The Elemental mapping of Fe, Cr and Te is also recorded which confirms the homogeneous distribution throughout the single crystal flake of $Fe_{1/5}CrTe_2$.

## Magnetization

To study the magnetic anisotropy, the DC magnetization measurements M(T) were performed on a bulk single crystal as a function of temperature along both principle crystallographic directions (ab-plane and c-axis). M(T) measurements were carried out with the DC field applied along the ab plane and c axis under the zero field cooled (ZFC) warming and field cooled (FC) warming protocols in temperature range of 2 K to 300K. A DC magnetic field of 500 Oe and 100 Oe were applied for field along the ab plane and c-axis respectively. Fig. 2 (a) show a steep rise in the DC magnetization of sample below 200 K. This indicates the presence of para to ferromagnetic transition. On decreasing the temperature, a gradual decrease of magnetization is observed. This can be due to competing magnetic interactions and anisotropy effects, which may be impacted by growth induced strain and Fe deficiency. After the initial decrease in magnetization, it becomes almost constant in temperature range 60 K to 30 K followed by an upturn below 25 K. The decrease in magnetization with decreasing temperature is a common feature in $Cr_xTe_y$ family compounds [40-44] due to spin canting from collinear to non-collinear structural transition. For the field applied parallel to the c-axis inset (ii) of Fig. 2(a) a similar trend of steep increase in magnetization was observed confirming the paramagnetic to ferromagnetic phase transition (i.e. $T_C \approx 182$ K) followed by a steep drop in magnetization near 50 K. This characteristic may indicate the existence of a secondary magnetic transition or a low-temperature reconfiguration of the magnetic state. The additional phase in lower temperature range may result from the anti-ferromagnetic ordering between Fe and Cr spins which favours an anti-parallel alignment [43, 45]. A comparable low-temperature anomaly has been observed in previous reports [24]. However, its microscopic origin has not been extensively studied. The nature of this transition would need to be clarified by additional microscopic investigations, such as neutron diffraction or local magnetic probes. The bulk single-crystal FCT exhibits strong magnetic anisotropy, as seen by the substantial differences in magnetization magnitude between the two crystallographic directions below the ferromagnetic transition temperature. The derivative of ZFC warming curve for H || c-axis crystallographic directions as shown in inset (i) of Fig. 2(a) confirms $T_C \approx 182$ K followed by additional phase transition $T_{C2} \approx 50$ K.

An additional magnetic properties like effective magnetic moment and Curie Weiss temperature is calculated from temperature dependent inverse susceptibility fit in temperature range 250 K to 300 K by using the Curie Weiss equation $\chi = C/(T – \theta_C)$ [46], where $\chi$ is the magnetic susceptibility, C and $\theta_C$ are Curie constant and Curie Weiss temperature respectively. By using the Curie Weiss equation in temperature range 250 K to 300 K, we calculate effective moment are $\approx 5.91$ $\mu_B$ per formula unit and $\approx 6.55$ $\mu_B$ per formula unit corresponding ab-plane and c-axis respectively. The calculated value of effective moment per formula unit lies in range of previously reported $FeCr_3Te_6$ [24] and $FeCr_2Te_4$ [45] sample results. The calculated value of Curie Weiss temperature are 205.48 K and 214.8 K for $\mu_oH$ || ab-plane and $\mu_oH$ || c-axis respectively. The positive value of Curie Weiss temperatures in both crystallographic directions are indicates ferromagnetic dominance characterises in our system.

$CrTe_2$ is the parent compound which exhibits a high Curie temperature of 320 K [20, 21]. As reported previously, the intercalation of Fe in this system results in the decrease of the Curie temperature (124 K) as observed in $FeCr_3Te_6$ [24]. According to previous reports, $FeCr_3Te_6$ is a Fe-intercalated derivative of $CrTe_2$ with an effective composition of $Fe_{1/3}CrTe_2$, in which Fe partially fills the interlayer sites between $CrTe_2$ layers [24]. In comparison to stoichiometric $FeCr_3Te_6$, the decreased Fe content in $Fe_{1/5}CrTe_2$ in the current investigation suggests a lower degree of Fe intercalation. The Curie temperature of $FeCr_3Te_6$ has been reported to be 124 K. Accordingly, the Curie temperature of $Fe_{0.6}Cr_3Te_6$ ($\approx 182$ K) falls between that of $CrTe_2$ [20, 21] and $FeCr_3Te_6$ [24]. This variation of $T_C$ with Fe content reflects the suppression of the ferromagnetic ordering temperature with Fe intercalation. This is probably due to the gradual alteration of magnetic interactions as the concentration of Fe rises.

In order to further describe the magnetic behaviour of this system, field dependent isothermal magnetization measurements were conducted in a field range of -7 T to +7 T as shown in Fig. 3(a). The steep increase in the magnetization followed by saturation at a very small applying external magnetic field ($\approx 0.5$ T) at several temperatures below $T_C$ indicates the ferromagnetic nature of bulk single crystal Fe$_{1/5}$CrTe$_2$. Inset (i) of Fig. 3(a) shows saturation magnetization at 2 K is 40.14 emu/gm at magnetic field 0.59 T with coercive field is 0.15 T. A decrease in the value of saturation magnetization is observed as the temperature increases. At 270 K, the field dependent magnetization depicts a linear behaviour which corresponding to the paramagnetic behaviour of the sample.

Inset (ii) of Fig. 3(b) shows temperature dependence of the saturated magnetization obtained by the linear extrapolation of the M-H behaviour at high fields. It shows that M$_S$ clearly follows a quadratic dependence with temperature up to 150 K rather than T$^{3/2}$ behaviour. Unlike the T$^{3/2}$ behaviour predicted for separate spin-wave excitations [39], the observed T$^2$ suppression of magnetization is typically linked to long-wavelength, low-frequency spin fluctuations or multiple excitations of interacting spin waves [36].

**Resistivity**

We now focus on examining electrical transport characteristics of the FCT single crystal. According to Matthiessen's rule different scattering mechanism can contribute to the longitudinal resistivity of a metallic sample. The total resistivity can be expressed as follows [47] $\rho_{xx}(T) = \rho_O + \rho_{e-m} + \rho_{e-p} + \rho_{e-e}$. Here, $\rho_O$, $\rho_{e-m}$, $\rho_{e-p}$ and $\rho_{e-e}$ corresponds to the residual resistivity, electron-magnon, electron-phonon, and electron-electron scattering respectively. The different scattering mechanisms tend to vary differently with temperature. For instance, the phonon term exhibits a linear T dependence at the high temperature, while both electron-electron and electron-magnon scattering term tend to vary quadratically with temperature. The residual resistivity is mainly caused by the defects, disorders, and imperfections in lattice periodicity. The resistivity as the function of temperature was measured in the temperature sweep from 2 K to 300 K. The resistivity increases with increasing temperature depicting the metallic nature of the sample. Near 182 K the resistivity behavior changes slope which indicates ferromagnetic to paramagnetic magnetic transition. This is consistent with the Curie temperature obtained from the temperature dependent magnetization behaviour. The residual resistivity ratio (RRR) of bulk single crystal sample is 1.55 which is comparable with the previous report [24]. The residual resistivity obtained for this sample is 0.3358 mΩ-cm. In order to identify the dominant scattering mechanisms, we fit resistivity vs. temperature in different temperature regime. Below the magnetic ordering temperature, the renormalized spin fluctuation theory [48] suggests that the electrical resistivity shows T$^2$ dependence for the itinerant ferromagnetic system [48]. To identify the dominating scattering mechanism, the resistivity data was fitted using two models: $\rho(T) = \rho_O + cT^2$ [48] and $\rho(T) = \rho_O + aT^{3/2} + bT^2$ [49]. As shown in Fig. (3), the incorporation of the $T^{3/2}$ temperature dependence provides a better fit to the resistivity data. Here $\rho_O$ is the residual resistivity a and b are temperature coefficients. The fitting yields $\rho_O$ = 0.3358 mΩ-cm, a = 2.65383E-5 mΩ-cmK$^{-1}$, and b = 1.20647E-6 mΩ-cmK$^{-2}$. The coefficient of $T^{3/2}$ is greater than the $T^2$ which indicates that the $T^{3/2}$ term dominates. The dominance of $T^{3/2}$ over $T^2$ reflects that the electrons are strongly coupled to the localized spins [49-51]. This emphasizes the significance of electron correlation effects in the FCT compound and prohibits a straightforward free-electron description. In FeCr$_2$Te$_4$ [50, 51] a member of the same family, has also been reported to exhibit similar behavior, indicating that the strong electron correlation may be a prevalent property in these materials. In paramagnetic region resistivity becomes almost linear with temperature indicating the dominance of the electron phonon scattering in the high temperature regime.

## Magnetoresistance

Owing to its possible topological characteristics [24], a thorough investigation of the magnetoresistance (MR) in FeCr$_3$Te$_6$ is necessary. Being a potential topological material [24], it is necessary to study the MR of such materials. Hence, we have performed isothermal MR at varied temperatures on bulk single crystal of FCT with dimensions (2.5×1.8×0.06) mm$^3$ along easy magnetization axis i.e. H || c-axis as shown in Fig. (4). To eliminate the transverse Hall resistivity contribution from MR data, the following equation was used

$$\rho_{xx}(H,T) = [\rho_{xx}(+H, T) + \rho_{xx}(-H, T)]/2 \qquad (1)$$

To calculate the MR from longitudinal resistivity data following equation was used:

$$MR = [\rho_{xx}(H, T) - \rho_{xx}(0, T)]/ \rho_{xx}(0, T), \qquad (2)$$

here $\rho_{xx}(0, T)$ and $\rho_{xx}(H, T)$ are the longitudinal resistivities at particular temperature in zero and finite applied magnetic field respectively. The isothermal MR reveals a linear non-saturating negative MR at low temperatures below $T_C$. The magnetic field can be divided into three different regimes for a systematic understanding of the MR behaviour: 0–0.15 T, 0.15–0.35 T, and 0.35–5 T. As shown in the inset of Fig. (4), below $T_C$, the MR becomes positive up to 0.15 T, reach a maximum value approx 0.25% for all measured temperature below $T_C$ and then declining sharply in the field range from 0.15 T to 0.35 T. A distinct change in slope occurs after 0.15 T resulting in a sharp decline in resistivity up to 0.35 T. After this, the MR becomes linearly negative and does not show saturation up to 5 T. Above the Curie temperature i.e., 190 K, MR is still negative but deviate from the linearity in the high field region as depicted in Fig. (4). Positive MR at lower field region can arise due to different mechanism like weak anti-localization (WAL) [52, 53], anisotropic MR and domain walls scattering [53, 54]. In this instance, the positive MR lasts up to 150 K, whereas weak anti-localization (WAL) effects are usually noticeable at low temperatures. WAL-induced positive MR is improbable at such high temperatures since the phase-coherence length is anticipated to decline. Consequently, the WAL effect cannot be responsible for the observed positive MR in our sample. Positive MR can also result from anisotropic MR (AMR) [54] when the magnetic field is applied parallel to the direction of the current. However, since the magnetic field is applied perpendicular to the current in the current setup, the AMR effect cannot be responsible for the observed positive MR [55, 56].

Ferromagnetic compounds have number of randomly orientated domain, thus electron are easily scattered from these domains and this results in a positive MR [56]. In the correlation study of MR and magnetization with the applied magnetic field parallel to the $c$ axis, as shown in stack Fig. (4), the MR of the bulk single crystal is initially positive and then becomes negative with increasing magnetic field. A clear change in slope is observed at a magnetic field of 0.35 T called critical field ($H_S$). The same magnetic field corresponds to Hs in the magnetization data, beyond which technical saturation is observed for all measured temperatures below the critical temperature. It is evident from Fig. (4) that below the Curie temperature, the MR shows a nearly linear field dependence. Negative MR in ferromagnetic systems can typically result from an applied magnetic field suppressing magnetic scattering [57]. Local magnetic moments scatter conduction electrons in heavy fermion compounds above the Kondo temperature scale [58]. The application of a magnetic field reduces this scattering, resulting in negative MR. Furthermore, simultaneous negative MR has also been frequently seen in ferromagnetic compounds, where the applied magnetic field reduces the inelastic scattering rate of conduction electrons by suppressing the fluctuations of the ferromagnetic order parameter [57, 59-61]. Such a mechanism can explain the behaviour observed in the present system. Moreover, the ferromagnetic correlations between localized spins, where the magnetic field gradually suppresses spin fluctuations and spin-disorder dispersion in a nearly proportionate manner, are typically responsible for the linear field dependence of MR [57, 59-61]. The long-range ferromagnetic order is weakened and thermal spin fluctuations take over as the temperature rises over $T_C$. This results in a departure from the linear MR behaviour and the appearance of nonlinear MR above $T_C$ since the suppression of spin disorder by the applied magnetic field no longer scales linearly with field [59]. Hence, the underlying magnetic phase transition is reflected in the clear shift in MR behaviour across $T_C$, which implies the involvement of different scattering mechanisms on either side of the transition.

**Anomalous Hall Effect**

We now present the isothermal Hall resistivity results of $Fe_{1/5}CrTe_2$ single crystal with the magnetic field applied along the c-axis and a DC current flowing in the ab-plane, as shown in the Fig. 5(a). A steep increase in the Hall resistivity is observed at low magnetic fields, followed by an almost linear dependence in the high-field regime, similar to the field dependent magnetization indicating the presence of a pronounced anomalous Hall effect [62]. As the temperature approaches $T_C$, the anomaly in the Hall resistivity gradually diminishes, and the response evolves toward a linear field dependence, further confirming the paramagnetic-to-ferromagnetic phase transition discussed earlier. To eliminate the contribution of the longitudinal resistivity arising from voltage-probe misalignment in the ferromagnetic compound, Hall resistivity was symmetrised with respect to positive and negative magnetic fields, using the following equation:

$$\rho_{xy}(H,T) = [\rho_{xy}(+H, T) - \rho_{xy}(-H, T)]/2. \qquad (3)$$

It is well known that the total Hall resistivity of a ferromagnetic compound consists of two main contributions, as described below

$$\rho^{Total}_{xy} = \rho^{O}_{xy} + \rho^{AHE}_{xy} = R_O B + 4\pi R_S M_S, \qquad (4)$$

here, $\rho^{Total}_{xy}$ is the total Hall resistivity, $\rho^{O}_{xy}$ and $\rho^{AHE}_{xy}$ are the ordinary hall resistivity and anomalous hall resistivity respectively. The ordinary Hall resistivity is a consequence of deflection in trajectory of the charge carriers due to the Lorentz force when they are subjected to the external magnetic field. Second term $\rho^{AHE}_{xy}$ namely anomalous hall resistivity occurs in ferromagnetic compound due to the interplay between scattering, magnetization and spin orbit coupling [25, 26]. $R_O$ and $R_S$ are the ordinary and anomalous Hall coefficient respectively. The high field $\rho^{Total}_{xy}$ data can be extrapolated to obtain the $R_O$ and $R_S$ values at different temperature. The slope obtained through the linear fitting of $\rho^{Total}_{xy}$ corresponds to $R_O$ which reflects the charge carrier density whereas the intercept corresponds to $\rho^{AHE}_{xy}$ respectively. The anomalous Hall coefficient $R_S$ can be obtained using the relation $R_S$ =

$\rho^{AHE}_{xy}/(4\pi M_S)$ [25, 26], where $M_S$ is the saturation magnetization. As shown in Fig. 5(a), the $\rho^{AHE}_{xy}$ shows a decreasing trend with increasing temperature, similar to the trend of $M_S$ with temperature. The maximum value of the Hall resistivity for $Fe_{1/5}CrTe_2$ is obtained as 2.8 μΩ-cm at a very small magnetic field (near 0.5 T) at temperature 2 K. Fig. 5(b) shows the variation of $R_O$ with temperature. The positive value of $R_O$ indicate that hole-type carriers dominate the charge transport process over the entire temperature range. The value of $R_O$ shows a considerable rise with increasing temperature which shows that the carrier density strength in this sample is sensitive to temperature. As shown in the Fig. 5(b), $R_S$ (T) is also positive and increases with increasing temperature. Notably, the magnitude of $R_S$ is approximately two orders of magnitude larger than $R_O$. This large difference indicates that the anomalous Hall resistivity strongly dominates over the ordinary Hall resistivity [25, 26]. Consequently, $Fe_{1/5}CrTe_2$ is a strong candidate for exhibiting a pronounced anomalous Hall effect.

Hall conductivity is calculated by the tensor relation $\sigma_{xy}^{AHE} = \rho_{xy}^{AHE}/[(\rho_{xy}^{AHE})^2 + (\rho_{xx})^2]$ [63]. Inset of Fig. 5(c) shows the decreasing trend of anomalous Hall conductivity with temperature. In ferromagnetic compounds, three primary mechanisms typically control the anomalous Hall effect (AHE): intrinsic contribution, extrinsic side-jump scattering, and extrinsic skew scattering [25, 26]. According to KL theory, the intrinsic anomalous Hall effect originates from band structure of materials. Charge carriers acquire an additional transverse velocity from berry curvature of occupied electronic band structure in momentum space known as anomalous hall velocity. The anomalous Hall resistivity's intrinsic contribution scales quadratically with the longitudinal resistivity [28, 29] (i.e., $\rho^{int.}_{xy} \sim \rho^2_{xx}$). Second mechanism is extrinsic skew scattering mechanism. According to Smit, skew scattering mechanism arises from the asymmetric scattering of spin polarized charge carrier from defects or impurities in presence of spin orbit coupling and unequal left and right scattering produced a net transverse current. Skew scattering contribution of anomalous Hall resistivity shows a linear dependence on the longitudinal resistivity [30, 31] i.e., $\rho^{sk}_{xy} \sim \rho_{xx}$. The third mechanism is extrinsic side-jump, which originates when spin-polarized charge carriers scatter from defects or impurities with significant spin-orbit coupling. According to Berger's theory, charge carriers with opposing spin orientations are deflected in opposing directions because they experience opposite effective electric fields [32]. As like intrinsic mechanism, side-jump contribution of the anomalous Hall resistivity shows a quadratic dependence on the longitudinal resistivity [32] (i.e., $\rho^{sj}_{xy} \sim \rho^2_{xx}$).

In order to determine the dominant mechanism of the origin of AHE, the following relation was used

$$\rho^{AHE}_{xy} = a(M)\,\rho_{xx} + b(M)\,\rho^2_{xx}, \qquad (5)$$

here, first term is known as the skew scattering contribution [30, 31] and second term represents the intrinsic [28, 29] or side jump contribution [32]. To study the dominance scattering contribution we should separate out the different scattering mechanism. In theory, it is highly challenging to distinguish between intrinsic and extrinsic contributions using measurements that depend on temperature or magnetic field on a single sample. This is because of the change in the temperature and magnetic field at the same time impact multiple parameters, including the magnetization M, the coefficients a(M) and b(M), and the longitudinal resistivity $\rho_{xx}$ [36, 38]. According to previous study the skew scattering contribution a(M) is usually proportional to M linearly [38]. Hence, the a(M) can be represented as a*M where a* is a constant. Accordingly, this constant a* in the skew scattering contribution a(M) can be obtained by plotting $\rho^{AHE}_{xy}/(M_S\rho_{xx})$ vs. $\rho_{xx}$. The plot $\rho^{AHE}_{xy}/(M_S\rho_{xx})$ vs. $\rho_{xx}$ displays a linear fit behaviour as shown in Fig. 5(d). The value of the parameter a, which measures the skew-scattering contribution, is obtained from the intercept of this linear fit. This value can be used to quantify the skew-scattering component of the anomalous Hall resistivity at various temperatures.

Furthermore, subtraction of this skew scattering contribution from the total anomalous Hall resistivity results in the obtaining of the intrinsic contribution to AHE. Hence, the skew scattering contribution and intrinsic contribution to anomalous Hall conductivity along with the total anomalous Hall conductivity has been depicted in Fig. 5(e). It can be seen clearly that the anomalous Hall conductivity is dominated by the extrinsic skew scattering contribution over the intrinsic mechanism. This behaviour is consistent with the previous report on $Fe_{1/3}CrTe_2$ [24]. To further comprehend the nature of the anomalous Hall response, the anomalous hall scale factor $S_H$ ($S_H = R_S/\rho^2_{xx} = \sigma_{xy}^{AHE}/M_S$) [64] was calculated as shown inset Fig. 5(e). $S_H$ remains almost constant with temperature below $T_C$ and weakly depends on magnetic field [64]. These $S_H$ values corroborate the validity of the scaling analysis because they fall within the usual range of 0.01-0.14 reported for ferromagnetic conductors [25]. Interestingly, over a wide temperature range, intrinsic anomalous Hall conductivity clearly shows a linear dependence on $M_S$ as shown in Fig. 5(f). This linear scaling suggests that the magnetic order parameter plays a major role in controlling the temperature evolution of the intrinsic Hall response. According to a spin-fluctuation framework that was used to interpret a comparable phenomenon in the ferromagnet $Mn_5Ge_3$, long-wavelength spin fluctuations lower the net magnetization without appreciably changing the underlying electronic structure [36]. Long-wavelength spin fluctuations may also be crucial in controlling the inherent anomalous Hall response in this system, as evidenced by the detection of a similar scaling pattern in $Fe_{1/5}CrTe_2$. This has been further confirmed by the linear scaling between $M_S$ versus $T^2$ in inset (ii) Fig. 2(b). Significantly, this result shows that disorder introduced by intercalation does not always suppress the underlying Berry-curvature-driven transport behaviour. It also shows that such a simplified intrinsic Hall scaling can continue even in an intercalated system with significant Fe-related disorder. This kind of behaviour has been previously observed in $Mn_5Ge_3$ system where it was attributed to the presence of long wavelength spin fluctuations [36].

## Topological Hall effect

In comparison to the anomalous Hall effect, which is caused by the momentum (k) space Berry curvature [65] and extrinsic scattering mechanisms, topological Hall originates from real-space berry curvature which is associated with the coupling of conduction electrons to chiral or nontrivial spin textures resulting in a fictitious magnetic field [65]. In CrTe based systems the emergence of a non-coplanar spin texture is primarily controlled by two important factors: (i) substantial anomalous Hall conductivity in the ferromagnetic regime [66] and (ii) competing FM and AFM interaction in the low temperature regime [66]. When the contribution from the topological Hall effect (THE) is taken into account, the total Hall resistivity can be expressed as the sum of the normal Hall resistivity, the anomalous Hall resistivity, and the topological Hall resistivity ($\rho^{THE}_{xy}$) [24].

$$\rho^{Total}_{xy} = \rho^{O}_{xy} + \rho^{AHE}_{xy} + \rho^{THE}_{xy} = R_O\mu_oH + S_H\rho^2_{xx}M_S + \rho^{THE}_{xy}. \qquad (6)$$

Where $R_O$ is the ordinary Hall coefficient and $S_H$ (i.e., $S_H = R_S/\rho^2_{xx}$) is anomalous Hall scaling factor which is a material-specific scale factor, weakly dependent of the magnetic field and temperature [64], at fixed temperature $\rho^2_{xx}$ is also approximately constant especially in ferromagnets exhibiting low magnitude of magnetoresistances. Generally, the topological Hall effect emerges in the low field regimes where the non-coplanar or chiral spin textures give rise to a finite scalar spin chirality. On the other hand, these nontrivial spin configurations are suppressed at sufficiently strong magnetic fields because the spins tend to align parallel to the applied field. The normal Hall resistivity ($\rho^O_{xy}$) and the anomalous Hall resistivity ($\rho^{AHE}_{xy}$) must be deducted from the overall Hall resistivity in order to derive the topological Hall contribution. The parameters $R_O$ and $S_H$ were determined using above equation. In particular, a linear fit to the high-field data of $\rho_{xy}/\mu_oH$ as a function of $\rho^2_{xx}M_S/\mu_oH$ produces $R_O$ from the intercept and $S_H$ from the slope, as illustrated in the inset of Fig. 6(a). The ordinary and anomalous Hall contributions are then computed using these parameters. Fig. 6 (a), at T = 20 K, shows the calculated $R_O\mu_oH + S_H\rho^2_{xx}M_S$ curve (black line) using the above fitted Ro and $S_H$ values, along with the total Hall resistivity $\rho^{Total}_{xy}$ (red line). The blue curve depicts the $\rho^{THE}_{xy}$ which has been extracted by subtracting the anomalous and normal Hall contributions from the total Hall resistivity. In high field region, it was observed that value of $R_O\mu_oH + S_H\rho^2_{xx}M_S$ fits quite well with experimental data because topological Hall resistivity expected zero in high field. The maximum of $\rho^{THE}_{xy}$ at 2 K is 0.611 μΩ-cm which 25.4 % of total anomalous Hall resistivity corresponding magnetic field 0.39 T. A similar analysis was carried out at different temperatures and it was observed that topological Hall resistivity tends to decrease with the increase in temperature with the peak position also shifting towards the lower field Fig. 6(b). The abrupt peak in $\rho^{THE}_{xy}$ is considered as a typical symbol of THE and can be clearly observed up to 150 K. Similar topological Hall signal evolution has been observed in $Cr_xTe_y$ systems and Fe-intercalated $CrTe_2$ compounds like $Fe_{1/3}CrTe_2$ [24], suggesting that this type of behaviour might be common to this class of materials. According to previous reports on comparable systems, the THE signal has been frequently linked to chiral spin textures, including skyrmion-like states [24]. However, microscopic probes like neutron scattering measurements or low-temperature Lorentz transmission electron microscopy are needed to directly establish these nontrivial spin patterns.

## Conclusion

In conclusion, we have thoroughly examined the magnetic and magnetotransport properties of the diluted Fe-intercalated van der Waals ferromagnet $Fe_{1/5}CrTe_2$. In comparison to the $Fe_{1/3}CrTe_2$ composition, the system shows an increased Curie transition temperature (≈ 180 K), highlighting the critical impact of Fe content in adjusting the magnetic exchange interactions within the $CrTe_2$ framework. The $M_S$ Vs $T^2$ scaling of magnetization indicates to the presence of long wavelength spin fluctuations. The temperature dependent resistivity in the low temperature regime follows a $T^{3/2}$ behaviour attributed to scattering from spin fluctuations and strong electron-spin coupling. At low temperatures, magneto-transport measurements show a linear, non-saturating negative MR, which is consistent with the suppression of spin fluctuations under applied magnetic fields. Analysis of the anomalous Hall effect shows that extrinsic skew-scattering contributions, which are probably associated with the Fe-related disorder, dominate the overall Hall contribution. However, throughout a broad temperature range, the intrinsic anomalous Hall conductivity shows a strong linear scaling with the saturation magnetization. Although, such linear scaling is generally linked to Berry-curvature-driven mechanisms, but in this case the persistence of such linear scaling over a wide temperature range in the presence of strong disorder indicates that the intrinsic Hall response is primarily controlled by the magnetic order parameter in a regime dominated by long-wavelength spin fluctuations. Furthermore, the detection of a finite topological Hall signal suggests that non-coplanar spin textures may arise in this system that necessitates additional microscopic studies.

## Acknowledgement

Manoj Lamba thanks University Grants Commission (UGC), India, for providing UGC-SRF fellowship. Satyabrata Patnaik acknowledges Anusandhan National Research Foundation (Grant No. ANRF/PAIR/2025/000029/PAIR-A) and DST Nano-Mission Project (CONCEPT, Grant No. 10/2019(G)/6) for consumables and equipment grants.

**Figure captions**

**Figure 1:** Rietveld refinment of polycrystalline FCT sample at room temperature. Inset (i) schematic view of the crystal structure of FCT. Inset (ii) out of plane X-ray diffraction pattern of single crystalline FCT at room temperature. Inset (iii) shows the single crystal Laue diffraction pattern which confirms its single crystalline nature. Inset (iv) EDX pattern for FCT. Inset (v) SEM image which confirms the layered morphology.

**Figure 2:** (a) Magnetization versus temperature curve at 500 Oe extenal DC magnetic field parallel to ab plane by swiping temperature from 2 K to 300 K. The black curve corrosponds to the ZFC warming data, red curve corresponds to the FC warming data respectively. The green curve shows inverse susceptibility with high temperature Curie-Wiess fit (orange). Inset (i) shows the derivative of ZFC warming for field applied parallel to c-axis (blue curve). Inset (ii) shows the DC magnetization when magnetic field applied along c-axis, where red curve represents FC warming, black corresponds to ZFC warming and green cureve shows the $\chi^{-1}$ with Curie Wiess fit in high temperature regime. (b) Magnetization versus applied external magnetic field swipe from +7 T to – 7 Talong c-axis at several constant temperatures. Inset (i) shows the zoomed picture of magnetization vs. magnetic field. Inset (ii) shows the $M_S$ vs. $T^2$ and the solid line is a linear fit up to 170 K.

**Figure 3:** Temperature dependent electrical resistivity of FCT. Inset shows the zoomed image for both blue (dash) and red (solid) fit for temperature dependence below Curie temperature.

**Figure 4:** Plots of magnetization and MR as a function of magnetic field at several constant temperatures (black dotted line shows the correlation between magnetization and magnetotransport properties). Inset figure shows the zoomed MR data in low field regime.

**Figure 5:** (a) Hall resistivity is plotted as a function of magnetic field at several constant temperatures for field applied parallel to c-axis. (b) Temperature dependent normal and anomalous Hall coefficient below Curie temperature. (c) Temperature dependent anomalous Hall resistivity and the inset showing the anomalous Hall conductivity. (d) Plot of $\rho^A_{xy}$ /($M_S\rho_{xx}$) vs. $\rho_{xx}$ red solid line being the linear fit below Curie temperature. (e) Temperature dependent anomalous skew scattering and intrinsic Hall conductivity show temperature dependent scaling factor. (f) The plot of intrinsic Hall conductivity $\rho^{AHE}_{xy-int.}$ vs. saturation magnetization $M_S$.

**Figure 6:** (a) Experimental observed, calculated and extracted topological Hall resistivity as a function of magnetic field at 20 K, inset shows the plot of $\rho^{AHE}_{xy}/\mu_o H$ vs. $M_S \rho^2_{xx}/\mu_o H$ is linear and value of $S_H$ is extracted. (b) Topological Hall resistivity as a function of magnetic field at several constant temperatures.

**Figure 1:**

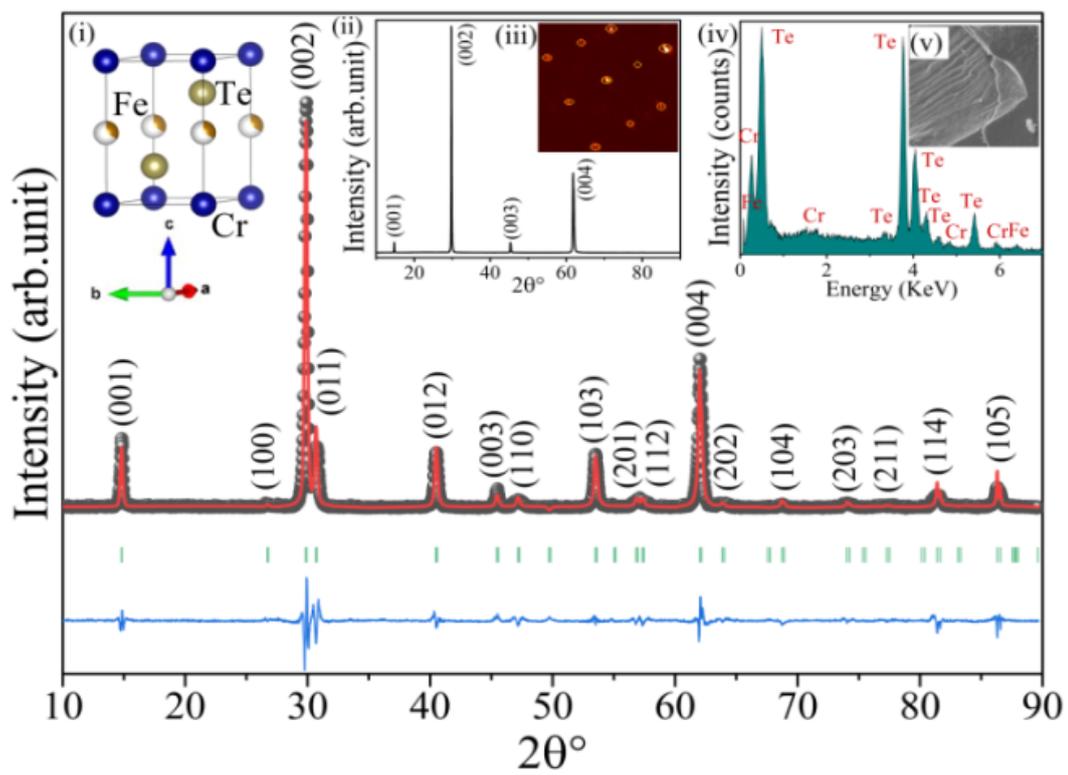

**Figure 2:**

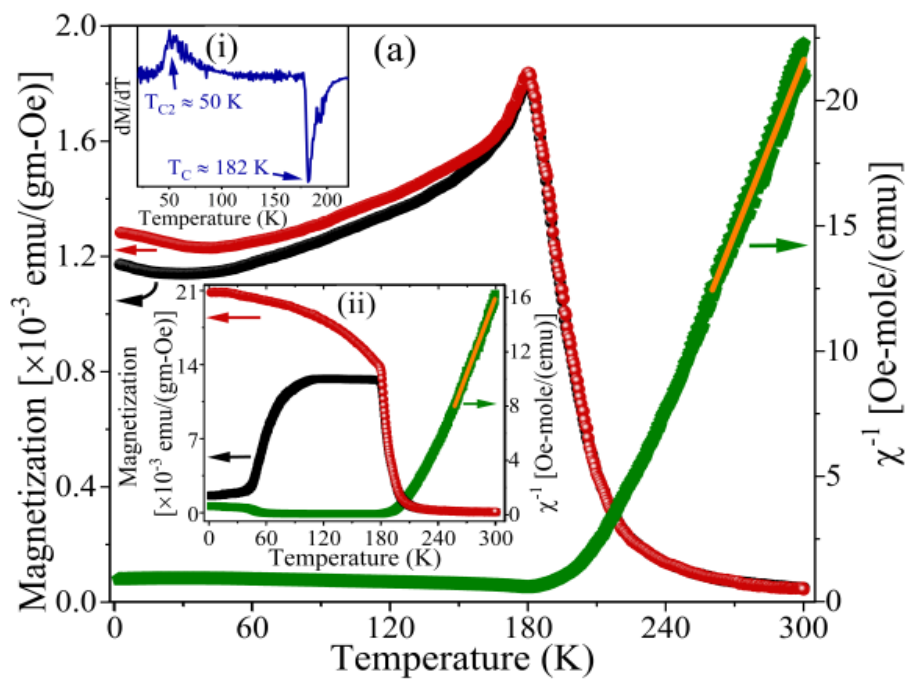

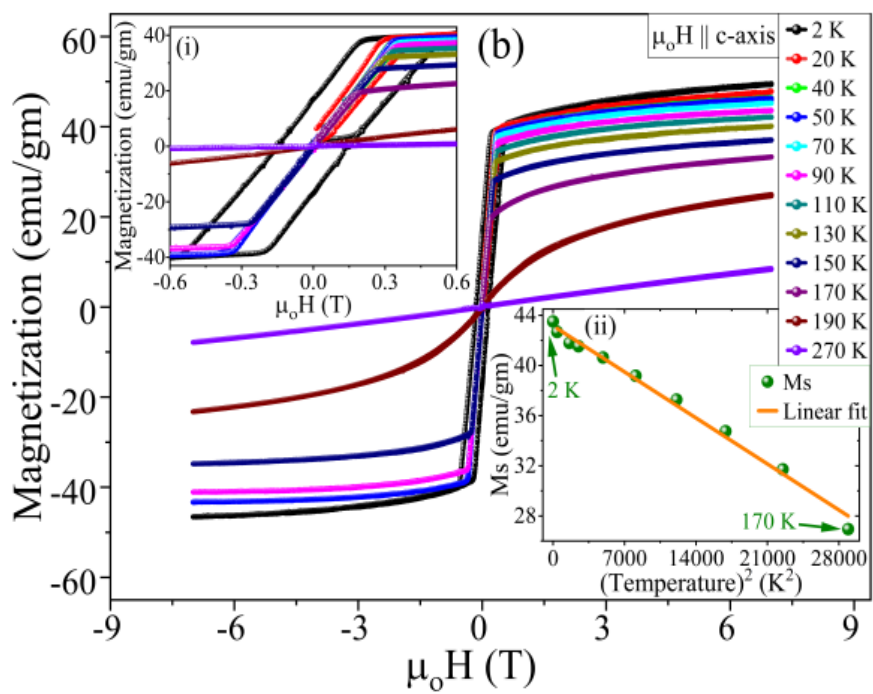

**Figure 3:**

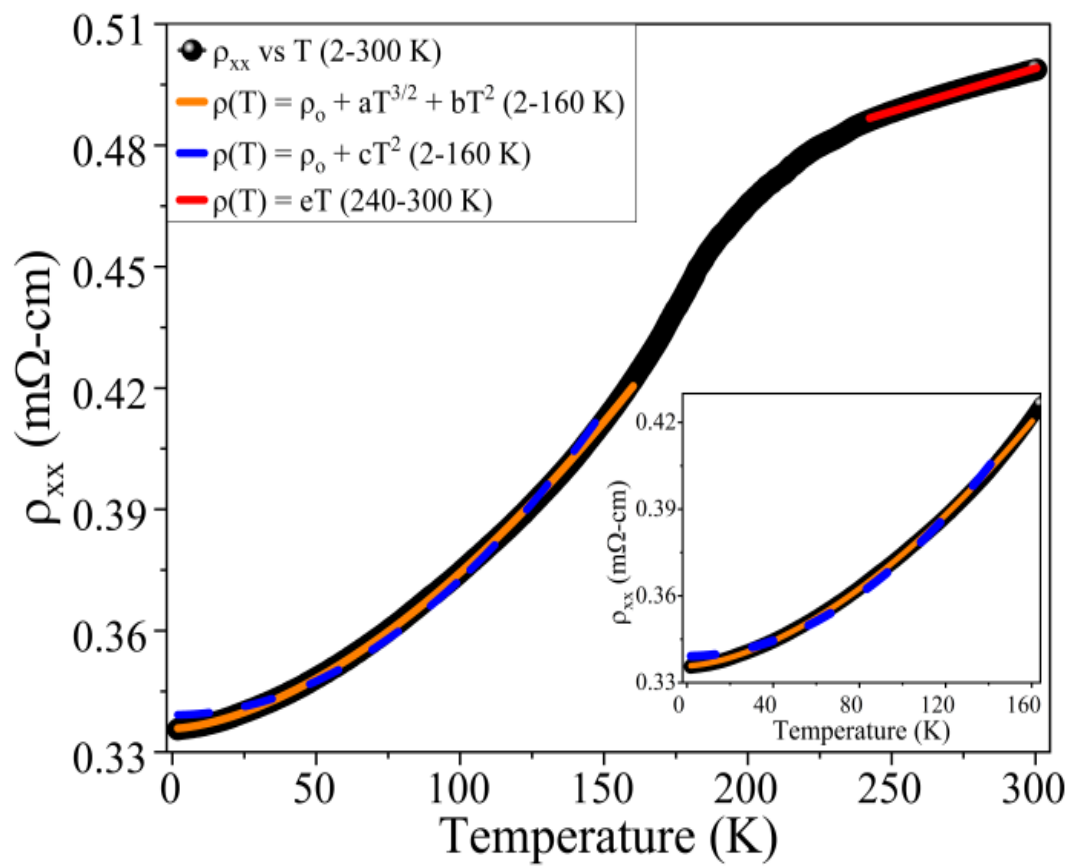

**Figure 4:**

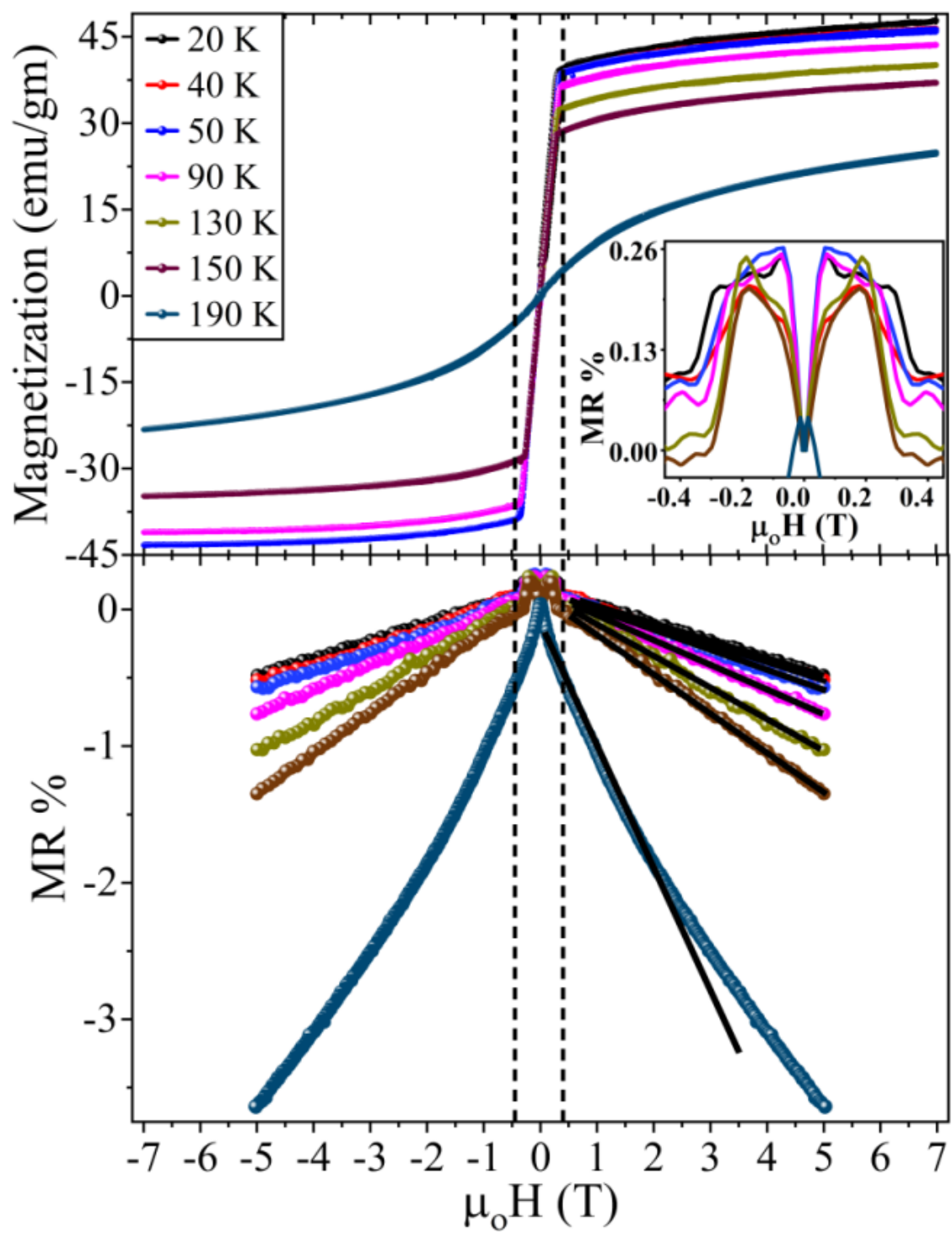

**Figure 5:**

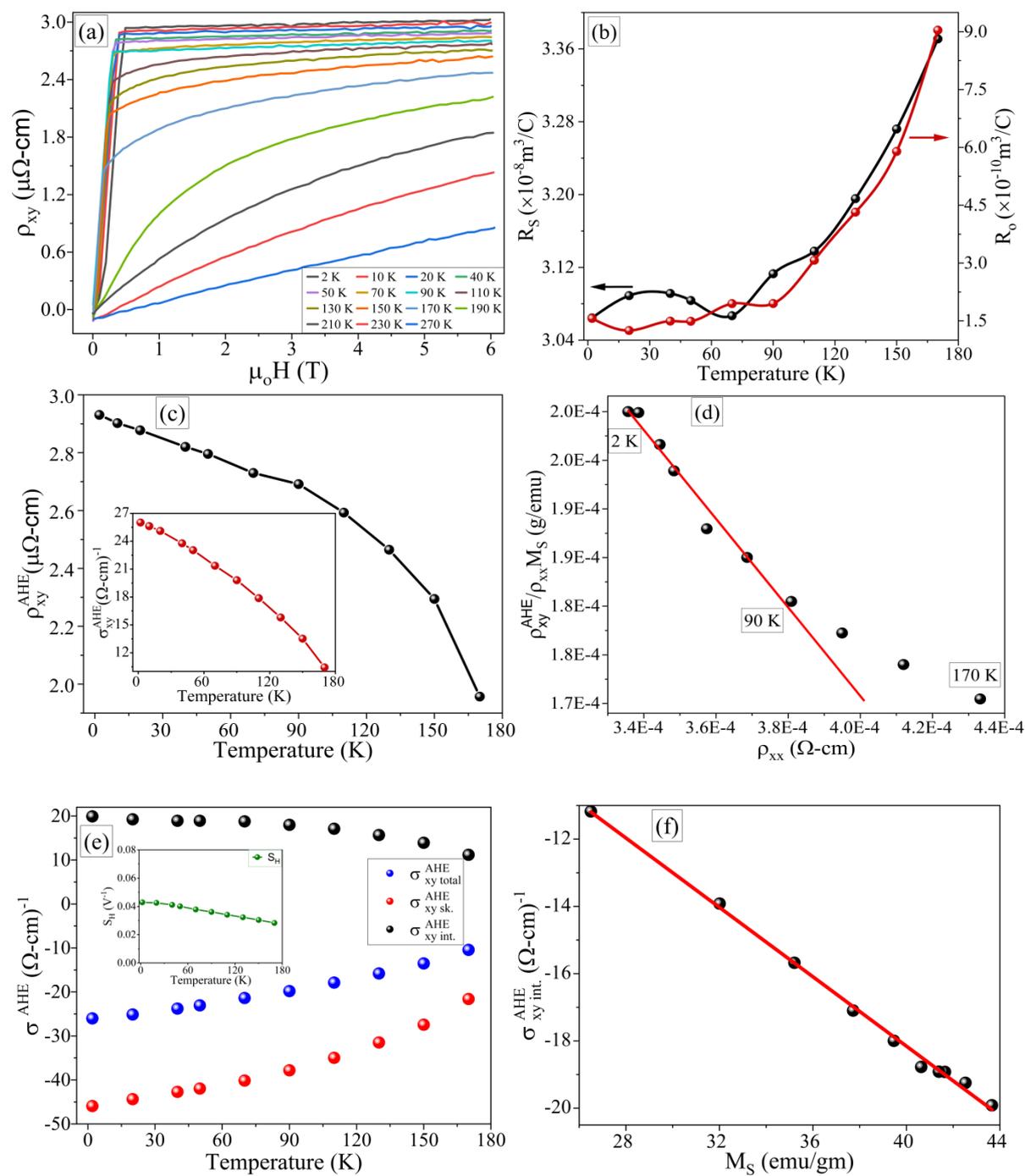

**Figure 6:**

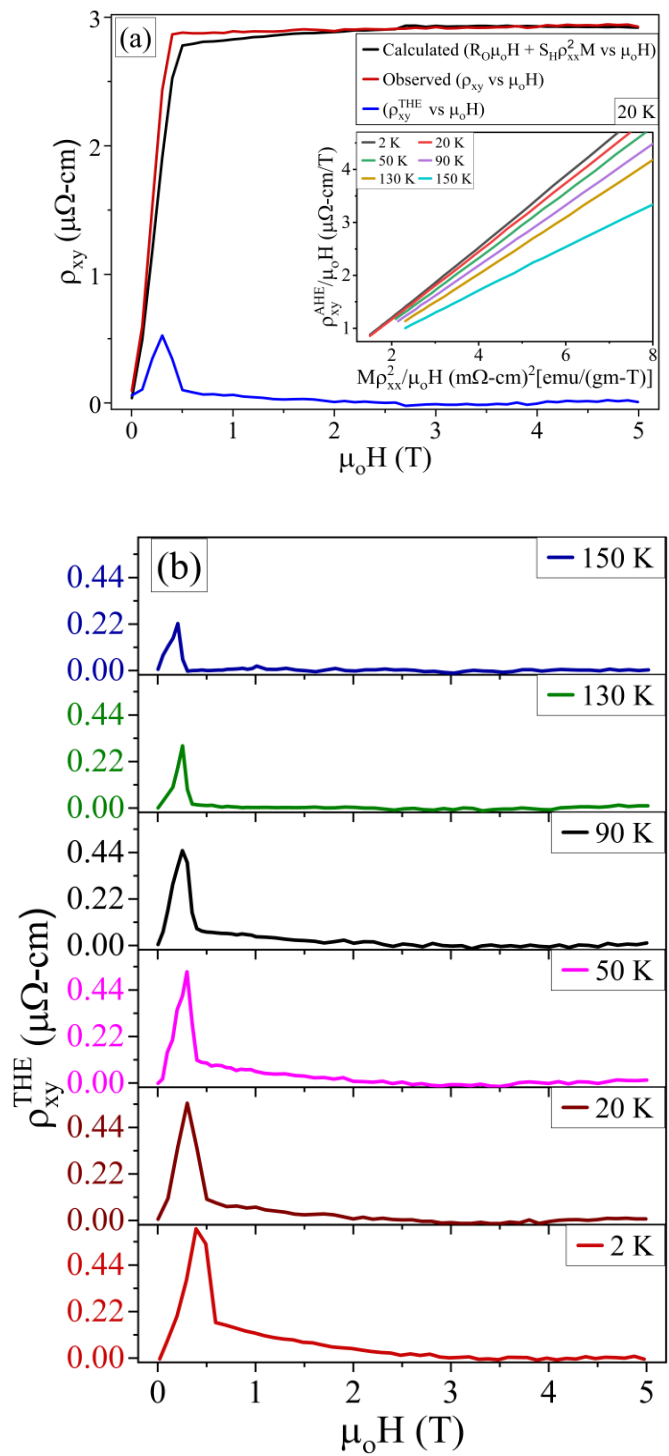